# Ultrafast charge separation in organic photovoltaics enhanced by charge delocalization and vibronically hot exciton dissociation

Hiroyuki Tamura[1] and Irene Burghardt[2]

[1]WPI-Advanced Institute for Material Research, Tohoku University, 2-1-1 Katahira, Aoba-ku, Sendai, 980-8577, Japan
[2]Institute of Physical and Theoretical Chemistry, Goethe University Frankfurt, Max-von-Laue-Str. 7, 60438 Frankfurt/Main, Germany

*Supporting Information Placeholder*

**ABSTRACT:** In organic photovoltaics, the mechanism by which free electrons and holes are generated overcoming the Coulomb attraction is a currently much debated topic. To elucidate this mechanism at a molecular level, we carried out a combined electronic structure and quantum dynamical analysis that captures the elementary events from the exciton dissociation to the free carrier generation at polymer/fullerene donor-acceptor heterojunctions. Our calculations show that experimentally observed efficient charge separations can be explained by a combination of two effects: First, the delocalization of charges which substantially reduces the Coulomb barrier, and second, the vibronically hot nature of the charge transfer state which promotes charge dissociation beyond the barrier. These effects facilitate an ultrafast charge separation even at low-band-offset heterojunctions.

Charge separation at donor-acceptor heterojunctions is a key process that determines the energy conversion efficiency of organic solar cells [1-16]. The photo-generated exciton is thought to primarily decay to a bound electron-hole pair localized at the donor-acceptor interface referred to as 'charge transfer (CT) state'[1-16] (Fig. 1). The donor and acceptor species are typically π-conjugated polymers, e.g., poly-3-hexylthiophene (P3HT), and fullerene derivatives, e.g., [6,6]-phenyl-$C_{61}$ butyric acid methyl ester (PCBM). The difference in the chemical potential between the anode and cathode induces an internal electric field of typically ~10 V/μm in the organic layer, favoring charge separation [1,11]. However, since the dielectric constant of organic materials is generally small ($\varepsilon_r$ = 3~4), the strong electron-hole Coulomb attraction stabilizes the interfacial CT state [1]. The potential barrier for the dissociation of point charges is typically ~0.4 eV, which is much higher than the thermal energy at room temperature (~0.026 eV). Moreover, free carrier formation can compete with the radiationless decay to the ground state, i.e., charge recombination, which reduces the internal quantum efficiency (IQE) [1,2].

How does the electron-hole pair separate into free carriers, and what are the key factors determining the efficiency of charge separation? To answer these questions, the following effects have been pinpointed in recent investigations: First, the delocalization of electron and hole [6,7,12-14,16] and second, the excess energy of the photo-generated exciton, entailing the so-called hot exciton dissociation mechanism [7-10]. However, no conclusive, general picture has been obtained to date.

In this study, we theoretically clarify how the interfacial CT state separates into free carriers at polymer/fullerene donor/acceptor interfaces (Fig. 1). A detailed microscopic analysis is carried out using quantum dynamical simulations with a parametrisation based on density functional theory (DFT) and time-dependent DFT (TDDFT) calculations. As detailed below, our analysis demonstrates the importance of charge delocalization in facilitating the generation of free carriers. We further propose a mechanism of charge separation mediated by vibronically hot CT states (Fig. 2). Our study shows how both effects work together to generate ultrafast and efficient charge separation.

The experimentally observed sub-picosecond charge separations [5,9] imply a coherent nature of the delocalized charges and excitons. In such systems, kinetic models such as Marcus theory [19] and the Onsager-Braun model [20] do not necessarily hold, such that non-perturbative quantum dynamical analysis is necessary. We explicitly account for electron-phonon (vibronic) couplings, which generally play an essential role in the charge and exciton transfers in organic semiconductors. To provide realistic parameters representing polymer/fullerene donor/acceptor heterojunctions such as P3HT/PCBM, we consider oligothiophene ($T_n$)/fullerene ($C_{60}$) interfaces as a model system (Fig. 1).

We employ a site-based model consisting of the exciton (XT), charge transfer (CT), and charge separated (CS) states, see Eq. (1) below (Methods section). The exciton coupling, the XT-CT coupling, the charge transfer integrals among the CT and CS states, as well as the vibronic couplings were determined by DFT and TDDFT calculations. The time-evolution of coherent vibronic wavepackets from the photo-generated exciton to free carriers is simulated using the multi-configuration time-dependent Hartree (MCTDH) method [27], analogously to our previous studies [21-26]. Such vibronic wavepackets delocalized over the molecular assembly (see Fig. 3(a)) correspond to a quantum superposition of the electronic-vibrational states of individual molecules.

Our previous DFT calculations [14] indicated that the potential barrier to charge separation decreases as the π-conjugation length of the donor molecule ($T_n$) increases, owing to the delocalization of the hole. The effective π-conjugation length of polymers is affected by the crystallinity at donor/acceptor interfaces, where annealing is known to enhance the crystallization [5,17].

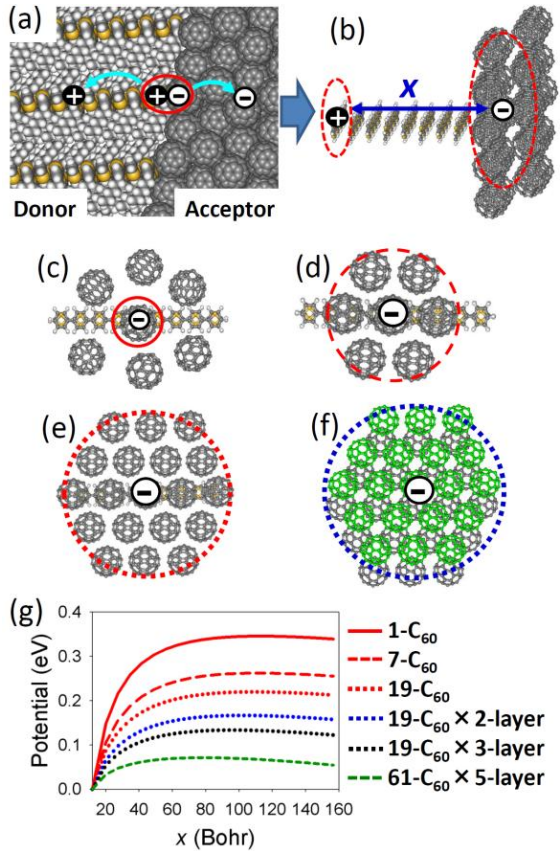

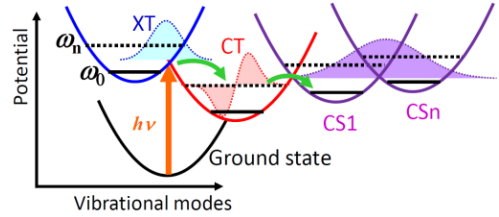

**Figure. 1**. (a) Schematic illustration of a crystalline domain of the P3HT/Fullerene donor/acceptor interface, considering a lamellar stacking of P3HT and a hexagonal close-packed cluster of fullerene. (b) Model systems consisting of π-stacked oligothiophenes ($T_n$: n=5~13) and a $C_{60}$ cluster. Here, the electron-hole distance, $x$, is defined as the center-to-center distance between a $T_n$ molecule and the $C_{60}$ molecule at the center of the first interface layer. (c) Model of disordered $C_{60}$ condensates, where one $C_{60}$ molecule interacts with $T_n$. Models of hexagonal close-packed $C_{60}$ clusters; (d) 7-$C_{60}$ × 1-layer, (e) 19-$C_{60}$ ×1-layer, and (f) 19-$C_{60}$ × 2-layer. (g) Potential curves for the charge separation considering a hole on $T_{13}$ and an electron on various sizes of $C_{60}$ clusters, based on the tight-binding model parametrized by DFT calculations, with a $C_{60}$-$C_{60}$ charge transfer integral of 0.05 eV, an electric field along the $x$ direction of 10 V/μm, and a dielectric constant of 4.

In this study, we further demonstrate that the modification of the potential barrier by electron delocalization in the fullerene condensates is of key importance. We consider the interface of hexagonal close-packed fullerene clusters as a model of the electron acceptors, and π-stacked $T_n$ moieties mimicking the lamellar structure of P3HT [18,28] in bulk heterojunctions (Fig. 1). The electron distribution on each fullerene is calculated based on a tight-binding model, where the fullerene on-site potential and the transfer integral have been determined by DFT calculations [14,26]. The lowest eigen state of the tight-binding Hamiltonian as a function of the electron-hole distance corresponds to the potential curve for the charge separation (Fig. 1). To investigate the effect of crystallinity, we calculate the electron-hole potential considering various sizes of fullerene clusters, where small vs. large clusters represent disordered vs. crystalline domains, respec-tively. Our calculations indicate that the potential barrier becomes shallower as the size of the fullerene cluster becomes larger (Fig. 1g), since the electron can delocalize over an increasing number of fullerene sites.

**Figure. 2.** Concept of charge separation mediated by vibronically hot CT states. Potential crossings of XT, CT, and CS states, together with the vibrational wavefunctions on the respective states. The solid and dashed black lines illustrate the vibrational ground ($\omega_0$) and excited ($\omega_n$) states, respectively.

To simulate the charge separation dynamics at a molecular level, the potential curve is mapped onto the on-site energies of the CS states consisting of an electron on the fullerene cluster and a hole on a $T_n$ moiety (Fig. 3a). Figures 3b and 3c show snapshots of the population distribution over the respective states during a 400 fs interval of our quantum dynamics simulations. The CT population rises within a few tens of femtoseconds (fs), and in turn the CS states are populated immediately (Fig. 3b-3e). Such ultrafast charge separations were indeed observed in the pump-probe experiments of polymer/fullerene photovoltaic systems [5,9]. As the inter-donor transfer integral increases, the charge separation becomes more efficient [see supporting information], where the hole can be delocalized over many molecules and can penetrate into the high potential sites, i.e., through-barrier tunneling can play a role. The CS states beyond the potential barrier (typically $CS_n$ with n > 10-15, (see Fig. 3a) are populated within 100~200 fs, which is much faster than the typical time scale of charge re-combination (i.e., a few hundred picoseconds) [2]. Hereafter, the sum of the CS populations beyond the barrier, i.e., the free carrier yield, is denoted as $\eta_{FC}$. The interfacial CT population exhibits a plateau after a few hundred fs, corresponding to carriers which remain trapped at the donor-acceptor interface and will eventually decay to the ground state on a longer time scale. That is, the major portion of $\eta_{FC}$ would be determined by the dynamics during the first few hundred fs.

While the role of charge delocalization in the electron-hole separation was previously conjectured [6,7,12-14,16], the present study is first to provide a microscopic insight into the ultrafast dynamics from the photogenerated exciton to free carriers, which can rationalize the observations by pump-probe experiments [3-5,9]. Our quantum dynamics calculations clearly indicate that the charge separation efficiency increases as the electron is more delocalized within the fullerene condensate, and as the π-conjugation length of the donor, i.e., the hole distribution length, increases (Fig. 4a). This is because the potential barrier is decreased by the charge delocalization (Fig. 1g). This trend is generally consistent with the experimental observations that indicate improvement of the IQE by increasing the regio-regularity of the donor polymer and by the annealing of polymer/fullerene blends [5] which enhances the crystallization. The experimentally reported barrierless charge separation [3,4] would be explained by the formation of such favorable domains in bulk heterojunctions.

In order to elucidate the role of vibronically hot CT states in the charge separation, we performed quantum dynamics simulations considering various values of the exciton-CT energy offset ($\Delta E_{XT\text{-}CT}$). Overall, the calculated $\eta_{FC}$ increases with increasing $\Delta E_{XT\text{-}CT}$ (Fig. 4b-4c). The excess energy of the exciton-CT transition induces vibrational excitations, such that the vibronically excited CT state can become resonant with CS states at energetically high potential sites (Fig. 2). On the timescale of a few hundred fs, the CT state is not yet equilibrated, and thus the vibronically hot CT states keep promoting the charge separation. In systems where $\Delta E_{XT\text{-}CT}$ is significantly larger than the barrier, $\eta_{FC}$ is not substantially increased by further increasing $\Delta E_{XT\text{-}CT}$ (Fig. 4b). A too large $\Delta E_{XT\text{-}CT}$ is unfavorable for exciton dissociation, since the potential crossing then tends to the inverted regime [19].

tor strengths than localized excitons on a single molecule. For illustration, we consider a coherent bright exciton delocalized over two sites which tends to result in a more efficient charge separation than the localized exciton (Fig. 4c). However, all the excess excitation energy is not necessarily exploited for the charge separation, because the bright exciton can rapidly decay to the lower-lying dark exciton.

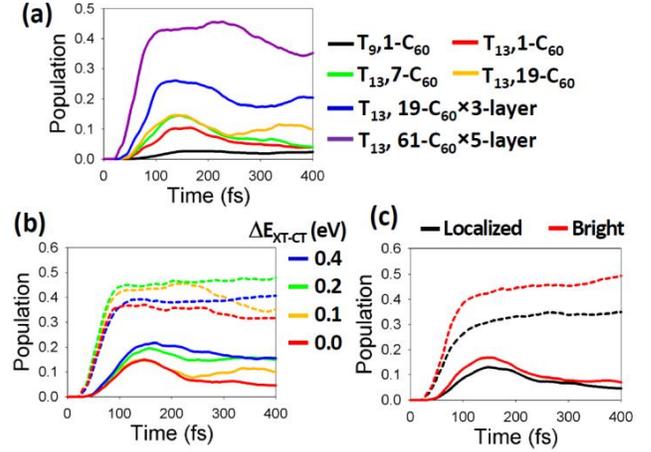

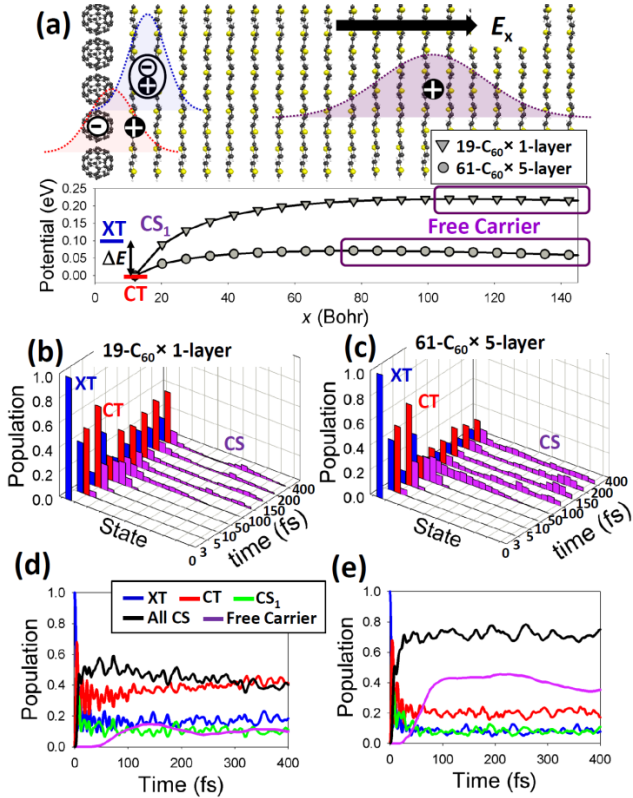

**Figure. 3**. (a) On-site energies of the respective CS states, together with schematic illustration of the photo-generated exciton, interfacial CT state, and delocalized polaron on the $T_n/C_{60}$ donor-acceptor model, where the $T_n$-$T_n$ distance is assumed to be 3.8 Å and an electric field, $E_x$, of 10 V/μm is applied perpendicular to the π-conjugation plane of the donor. Snapshots of the population distribution on the respective states during the quantum dynamics calculations for (b) $T_{13}$/[19-$C_{60}$×1-layer] and (c) $T_{13}$/[61-$C_{60}$×5-layer] donor/acceptor models. Population plots for these calculations; (d) $T_{13}$/[19-$C_{60}$×1-layer] and (e) $T_{13}$/[61-$C_{60}$×5-layer]. The sum of CS populations beyond the potential barrier is defined as the free carrier population (see panel (a)). The XT-CT coupling of 0.2 eV and the $T_n$-$T_n$ charge transfer integral of 0.12 eV are considered for all the calculations.

**Figure. 4**. (a) Free carrier yield $\eta_{FC}$ from the quantum dynamics calculations considering various sizes of $C_{60}$ clusters and π-conjugation lengths of $T_n$, where $\Delta E_{XT\text{-}CT}$ is set to 0.1 eV. (b) $\eta_{FC}$ for various values of $\Delta E_{XT\text{-}CT}$. (c) $\eta_{FC}$ generated from the bright exciton delocalized over two donor molecules, where $\Delta E_{XT\text{-}CT}$ for a single exciton is set to 0.1 eV and the exciton coupling is 0.15 eV. The solid and dashed lines indicate $\eta_{FC}$ using the potentials for the $T_{13}$/[19-$C_{60}$×1-layer] and $T_{13}$/[61-$C_{60}$×5-layer] models, respectively. The XT-CT coupling of 0.2 eV and the $T_n$-$T_n$ charge transfer integral of 0.12 eV are considered for all calculations.

Another pathway to charge separation was recently proposed [9], involving higher excited states of the donor molecule. Our DFT and quantum dynamics calculations suggest that this *electronically* hot exciton dissociation plays a minor role in the overall quantum efficiency, because of the low oscillator strength of the higher excitations and the small hole transfer integral between the highest and lower occupied molecular orbitals of the π-stacked $T_n$ (see supporting information).

In conclusion, our calculations revealed two main factors that can significantly enhance the free carrier generation. First, the lowering of the Coulomb barrier due to charge delocalization is an indispensable condition for the efficient charge separation. The charge delocalization can be realized in actual organic solar cells by enhancing the crystallization [5,17]. The barrierless charge separation observed in some experiments [3,4] can be rationalized by a small potential barrier due to charge delocalization. Second, vibronically hot CT states can enhance the ultrafast charge separation, where $\eta_{FC}$ can increase substantially as the excess energy increases, as reported in Refs. [7,9]. The excess energy varies with the exciton-CT offset ($\Delta E_{XT\text{-}CT}$) of the donor-acceptor heterojunctions as well as the delocalization of exciton. We found that the vibronically hot CT dissociation is particularly effective when $\Delta E_{XT\text{-}CT}$ is comparable to the Coulomb barrier. Too much excess energy is not effective for further improvement of $\eta_{FC}$, such that the use of large-$\Delta E_{XT\text{-}CT}$ materials is not necessarily a strategy for optimization.

Besides $\Delta E_{XT\text{-}CT}$, the delocalization of the photo-generated exciton on the π-stacked H-aggregate contributes to modifying the excess energy. Bright excitons delocalized over H-aggregates generally possess higher excitation energies and stronger oscilla-

Organic solar cells simultaneously demand large open circuit voltage and long-wavelength absorption for improving the energy conversion efficiency. In this context, efficient charge separation with a small excess energy, i.e., using low-$\Delta E_{XT-CT}$ materials, is favorable. While these requirements are incompatible with the presence of a high Coulomb barrier, the barrier can be reduced substantially by charge delocalization as shown in the present Communication. As a result, free carriers can be generated on an ultrafast time scale, without involving higher electronic excitations of the donor species. The present picture is generally applicable for the ultrafast dynamics of photo-induced charge separations in a broad range of heterojunction systems.

**Methods.** The excited states of $T_n/C_{60}$ systems were calculated using the long-range corrected time-dependent DFT (LC-TDDFT) [29] with the BLYP functional, where we consider the standard range separation parameter. The double-$\zeta$ basis set with $d$-function is employed with the SBKJC pseudopotential. The GAMESS code [30] is used for all of the DFT calculations.

Quantum dynamics calculations of the charge separation were carried out using the MCTDH method [27], based on a linear vibronic coupling model in a site-based diabatic representation. The Hamiltonian takes the following form,

$H = h_{XT}(x) |XT><XT| + h_{CT}(x) |CT><CT|$
$+ \Sigma_n h_{CSn}(x) | CS_n><CS_n| + \gamma (|XT><CT| + |CT><XT|)$
$+ t (|CT><CS_1| + |CS_1><CT|) + \Sigma_{nn'} t |CS_n><CS_{n'}|$ (1)

with the individual diagonal contributions $h_\xi(x) = \Sigma_i \omega_i/2 (x_i^2 + p_i^2) + \Sigma_i \kappa_i^{(\xi)} x_i + \varepsilon^{(\xi)}$, where $\xi = (XT, CT, CS_n)$ denotes the exciton, the charge transfer state, and the charge separated states, respectively; $x = \{x_i\}$ collectively denotes a set of intra-molecular phonon modes; $h_\xi(x)$ are the Hamiltonians of these harmonic modes in the respective states, and $\omega_i$ and $p_i$ are the frequency and momentum of the vibrational modes, respectively. Further, $\varepsilon^{(\xi)}$ are the respective on-site potentials. The XT-CT coupling $\gamma$, the charge transfer integral $t$, and the spectral density of vibronic couplings $\kappa_i$, were determined by DFT and TDDFT calculations [14,24-26]. The XT-CT offset is defined as $\Delta E_{XT-CT} = \varepsilon_{XT} - \varepsilon_{CT}$. In this study, we consider a reasonable range of $\Delta E_{XT-CT}$ based on the LC-TDDFT calculations for $T_n/C_{60}$ complexes of different π-conjugation lengths [14].

The vibrational modes were reduced to a limited number of effective modes, which reproduce the short-time dynamics and the reorganization energy of the whole system [21-26]. We considered a total number of effective modes of 110 for the present system, i.e., 8 modes of $C_{60}$, 8 modes of the first $T_n$, 7 modes of the second and third $T_n$, and 5 modes of the fourth to nineteenth $T_n$.

## ASSOCIATED CONTENT

### Supporting Information
Details of the calculations of electron-hole potentials, DFT parametrizations, relaxation of the bright exciton, and the dynamics of electronically hot exciton dissociation. This material is available free of charge via the Internet at http://pubs.acs.org.

## AUTHOR INFORMATION
### Corresponding Author
hiroyuki@wpi-aimr.tohoku.ac.jp

## ACKNOWLEDGMENT
This study is supported by the Grants-in-Aid for scientific research (C) from JSPS, Japan, as well as the Japanese-German NAKAMA funds. AIMR, Tohoku University, is supported by WPI program, MEXT, Japan.

## REFERENCES

[1] Deibel, C.; Strobel, T.; Dyakonov, V. *Adv. Mater*. **2010,** *22*, 4097–4111.
[2] Hwang, I. W.; Soci, C.; Moses, D.; Zhu, Z.; Waller, D.; Gaudiana, R.; Brabec, C. J.; Heeger, A. J. *Adv. Mater*. **2007,** *19*, 2307–2312.
[3] Pensack, R. D; Asbury, J. B. *J. Am. Chem. Soc*. **2009,** *131*, 15986-15987.
[4] Lee, J.; Vandewal K.; Yost S. R.; Bahlke M. E.; Goris L.; Baldo M.A.; Manca J.V.; Van Voorhis T. *J. Am Chem Soc*. **2010,** *132*, 11878-11880.
[5] Guo, J.; Ohkita, H.; Benten, H.; Ito, S. *J. Am. Chem. Soc.* **2010,** *132*, 6154–6164.
[6] Bakulin A. A.; Rao A.; Pavelyev V. G.; van Loosdrecht P. H. M.; Pshenichnikov M. S.; Niedzialek D.; Cornil J.; Beljonne D.; Friend R. H. *Science* **2012,** *335*, 1340-1344.
[7] Dimitrov, S. D.; Bakulin, A. A.; Nielsen, C. B.; Schroeder, B. C.; Du, J.; Bronstein, H.; McCulloch, I.; Friend, R. H.; Durrant J. R. *J. Am. Chem. Soc*. **2012,** *134*, 18189-18192.
[8] Jailaubekov A. E.; Willard A. P.; Tritsch J. R.; Chan W.-L.; Sai N.; Gearba R.; Kaake L. G.; Williams K. J.; Leung K.; Rossky P. J.; Zhu X-Y. *Nat. Mater*. **2013,** *12*, 66-73.
[9] Grancini, G.; Maiuri M.; Fazzi1 D.; Petrozza1 A.; Egelhaaf H-J.; Brida D.; Cerullo G. Lanzani G. *Nat. Mater*. **2013,** *12*, 29-33.
[10] (a) Armin, A., Zhang, Y., Burn, P. L., Meredith, P. & Pivrikas, A.; (b) Scharber, M. Measuring internal quantum efficiency to demonstrate hot exciton dissociation. *Nat. Mater.* (Correspondence on Ref. 9) **2013,** *12*, 593-594.
[11] Yuan Y.; Reece T. J.; Sharma P.; Poddar S.; Ducharme S.; Gruverman A.; Yang Y.; Huang J. *Nat. Mater*. **2011,** *10*, 296-302.
[12] Nayak P. K.; Narasimhan K. L.; Cahen D. *J. Phys. Chem. Lett.* **2013,** *4*, 1707-1717.
[13] Deibel, C.; Strobel, T.; Dyakonov, V. *Phys. Rev. Lett*. **2009,** *103*, 036402-1-4.
[14] Tamura, H.; Burghardt, I. *J. Phys. Chem. C* **2013,** *117,* 15020–15025.
[15] Rozzi, C. A.; Falke, S. M.; Spallanzani, N.; Rubio, A.; Molinari, E.; Brida, D.; Maiuri, M.; Cerullo, G.; Schramm, H.; Christoffers, J.; Lienau, C. *Nat. Commun*. **2013,** *4*, 1602-1-7.
[16] Caruso, D.; Troisi, A. *Proc. Natl. Acad. Sci. USA*. **2012,** *109*, 13498-13502.
[17] Savenije T. J.; Kroeze J. E.; Yang X.; Loos J. *Adv. Funct. Mater*. **2005,** *15*, 1260–1266.
[18] Kim, Y.; Cook, S.; Tuladhar, S. M.; Choulis, S. A.; Nelson, J.; Durrant, J. R.; Bradley, D. D. C.; Giles, M.; McCulloch, I.; Ha, C.-S.; Ree, M. *Nat. Mater.* **2006,** *5*, 197-203.
[19] Marcus, R. A. *Rev. Mod. Phys*. **1993,** *65*, 599-610.
[20] Braun, C. L. *J. Chem. Phys*. **1984,** *80*, 4157-4161.
[21] Tamura, H; Ramon, J. G. S.; Bittner, E. R.; Burghardt, I. *Phys. Rev. Lett.* **2008,** *100*, 107402-1-4.
[22] Tamura, H.; Bittner, E. R.; Burghardt, I. *J. Phys. Chem. B* **2008,** *112*, 495-506.
[23] Tamura, H.; Bittner, E. R.; Burghardt, I. *J. Chem. Phys*. **2007,** *127*, 034706-1-18.
[24] Tamura, H.; Burghardt, I. Tsukada, M. *J. Phys. Chem. C* **2011,** *115*, 10205-10210.
[25] Tamura H.; Martinazzo R.; Ruckenbauer M.; Burghardt I. *J. Chem. Phys*. **2012,** *137*, 22A540-1-8.
[26] Tamura H.; Tsukada M. *Phys. Rev. B* **2012,** *85*, 054301-1-8.
[27] Beck, M. H.; Jäckle, A.; Worth, G. A.; Meyer, H. D. *Phys. Rep.* **2000,** *324*, 1-105.
[28] Dag. S.; Wang, L. W. *J. Phys. Chem. B* **2010,** *114,* 5997–6000.
[29] Towada, Y.; Tsuneda, T.; Yanagisawa, S.; Yanai, Y.; Hirao, K. *J. Chem. Phys.* **2004,** *120*, 8425-8433.
[30] Schmidt, M. W.; Baldridge, K. K.; Boatz, J. A.; Elbert, S. T.; Gordon, M. S.; Jensen, J. H.; Koseki, S.; Matsunaga, N.; Nguyen, K. A.; Su, S.; Windus, T. L.; Dupuis, M.; Montgomery, J. A. *J. Comput. Chem.*, **1993,** *14*, 1347-1363.